\documentclass[aps,pra,twocolumn,groupedaddress,showpacs]{revtex4}
\usepackage{natbib}
\usepackage[latin1]{inputenc}
\usepackage[dvips]{graphicx}
\usepackage{amsmath,amsbsy,amsfonts,amssymb}
\usepackage{bm}

\newcommand{\abs}[1]{\left\vert#1\right\vert}
\newcommand{\half}{\ensuremath{\frac{1}{2}}}
\newcommand{\RE}{\operatorname{Re}}
\newcommand{\IM}{\operatorname{Im}}
\newcommand{\ket}[1]{\ensuremath{\left|#1\right>}}
\newcommand{\cafp}{\ensuremath{^{40}\text{Ca}^{+}}}
\newcommand{\srp}{\ensuremath{^{88}\text{Sr}^{+}}}
\newcommand{\bap}{\ensuremath{^{138}\text{Ba}^{+}}}
\newcommand{\omsub}[1]{\ensuremath{\omega_{#1}}}
\newcommand{\omz}{\ensuremath{\omega_{z}}}
\newcommand{\state}[3]{\ensuremath{\,^{#1}{#2}_{#3}}}
\newcommand{\unit}[1]{\ensuremath{\,\mathrm{{#1}}}}
\newcommand{\down}{\ensuremath{\ket{\downarrow}}}
\newcommand{\downdown}{\ensuremath{\ket{\downarrow}\ket{\downarrow}}}
\newcommand{\up}{\ensuremath{\ket{\uparrow}}}
\newcommand{\upup}{\ensuremath{\ket{\uparrow}\ket{\uparrow}}}
\newcommand{\downup}{\ensuremath{\ket{\downarrow}\ket{\uparrow}}}
\newcommand{\updown}{\ensuremath{\ket{\uparrow}\ket{\downarrow}}}

\begin{document}
\title{Geometric quantum gate for trapped ions based on optical dipole forces induced by Gaussian laser beams}
\author{Peter Staanum}
\author{Michael Drewsen}
\author{Klaus M\o lmer}
\affiliation{QUANTOP - Danish National Research Foundation Center
for Quantum Optics, Department of Physics and Astronomy,
University of Aarhus, DK-8000 Aarhus C, Denmark.}
\date{\today}

\begin{abstract}
We present an implementation of quantum logic gates via internal
state dependent displacements of ions in a linear Paul trap caused
by optical dipole forces. Based on a general quantum analysis of
the system dynamics we consider specific implementations with
alkaline earth ions. For experimentally realistic parameters gate
infidelities as low as $10^{-4}$ can be obtained.
\end{abstract}
 \maketitle

\section{Introduction}\label{sec:intro}

Today, one of the most promising physical systems for an
experimental realization of quantum computation is a string of
trapped ions~\cite{Cirac-Zoller}. In such a realization, a quantum
bit (qubit) is represented by two internal states of an ion on
which quantum logic operations can be performed through laser-ion
interactions. Any quantum logic operation can, e.g., be composed
of single-qubit operations and Controlled-NOT (CNOT) gates between
any two ions in the string~\cite{DiVincenzo-universal}. While
single-qubit gates are relatively simple to perform, the CNOT and
equivalent two-ion gates are more demanding and usually also slow
as compared to single-ion gates. Only very recently such two-ion
gates have been demonstrated
experimentally~\cite{Schmidt-Kaler-CNOT, Leibfried-CNOT}. One
class of two-ion gates is the geometric quantum gates, which can
be relatively fast as compared to other two-ion gates and
furthermore allows a high fidelity since the internal states of
the ions are not directly involved in the gate
operation~\cite{Leibfried-CNOT,Anders-general}.

In this paper we present a proposal for an experimental
realization of a geometric quantum gate. The idea underlying our
gate proposal is that internal state dependent optical dipole
forces applied to a pair of ions will displace their equilibrium
positions in the trap in proportion to the forces acting on them.
Their relative distance is thus modified, and the associated shift
of their Coulomb potential energy gives rise to a complex phase
factor which depends on the internal states of both ions, and
which hence provides a non-trivial two-qubit gate.

The paper is organized as follows. In Sec.~\ref{sec:theory} we
present a full quantum analysis of the dynamics of the system to
obtain the correct values for the phase factors and to properly
assess the effect on the atomic motion in the trap. In
Sec.~\ref{sec:implementation}, we consider specific
implementations in alkaline earth ions. Assuming a Gaussian
intensity profile of the applied laser beam, we find in
Sec.~\ref{sec:Gaussian beams} experimentally relevant parameters
for an implementation of the gate proposal using \cafp\ or \bap\
ions and discuss relevant error sources. Finally, in
Sec.~\ref{sec:conclusion}, we give a discussion and a conclusion.

\section{Theory}\label{sec:theory}
We consider a pair of ions confined to the trap axis of a linear
Paul trap, i.e., the axis where the rf-field
vanishes~\cite{Ghosh}. A laser beam which is far-off resonant with
all internal transitions in the ions and propagates perpendicular
to the trap axis induces an optical dipole force. This force can
be described by a potential $U_{dip}(z,\alpha )$ which depends on
the position variable $z$ along the trap axis, and on the internal
state label $\alpha$ taking one of two possible values,
represented by $\uparrow $ and $\downarrow $ in the following. The
potential is controlled by the intensity, waist, polarization, and
wavelength of the laser beam~\cite{Grimm-opt}.

For an ion string illuminated by a dipole force inducing beam its
total potential energy, $U$, is the sum of the dipole potential of
each ion, the potential energy of the ions due to the trap and due
to their mutual Coulomb repulsion. Specifically for two
singly-charged ions of mass $m$, we have
\begin{align}\label{eq6:pot-energy}
  U&(z_{1},z_{2},\alpha_{1},\alpha_{2})=\half
  m\omz^{2}(z_{1}^{2}+z_{2}^{2})+\frac{e^{2}}{4\pi\epsilon_{0}|z_{2}-z_{1}|}\\\nonumber
  &+U_{dip}(z_{1,eq},\alpha_{1})-F_{dip}(z_{1,eq},\alpha_{1})(z_{1}-z_{1,eq})\\ \nonumber
&+U_{dip}(z_{2,eq},\alpha_{2})-F_{dip}(z_{2,eq},\alpha_{2})(z_{2}-z_{2,eq}),
\end{align}
where \omz\ is the single ion oscillation frequency in the trap
and where a linear expansion of the dipole potential around the
equilibrium positions $z_{i,eq}$ ($i=1,2$) has been applied.
$F_{dip}(z_{i,eq},\alpha_i)=-\partial U_{dip}(z,\alpha_i)/\partial
z(z=z_{i,eq})$ is the optical dipole force exerted on the $i$'th
ion.

Rather than considering a CNOT gate, we shall in the following
focus on implementation of the equivalent Controlled-Z two-qubit
gate~\cite{Nielsen-Chuang}. A Controlled-Z gate applied to a
superposition of the product states \downdown, \downup, \updown\
and \upup\ changes the sign of the \upup\ term (i.e., the phase of
this state is changed by $\pi$), while leaving the others
unchanged. The implementation of a geometric Controlled-Z gate
relies on the fact that the dipole force is internal state
dependent. The origin of this dependence will be described in the
next section. As mentioned above, the idea is that the dipole
force displaces the ions away from their equilibrium positions,
and the associated change in Coulomb energy gives rise to
phase-shifts which depend on the internal states of the ions. By
choosing a suitable temporal and spatial profile of the dipole
force inducing beam, the obtained phase shifts of the above
mentioned four product states can be made equivalent to a
Controlled-Z gate.

\begin{figure}[!htbp]
\centering\includegraphics[width=0.9\linewidth]{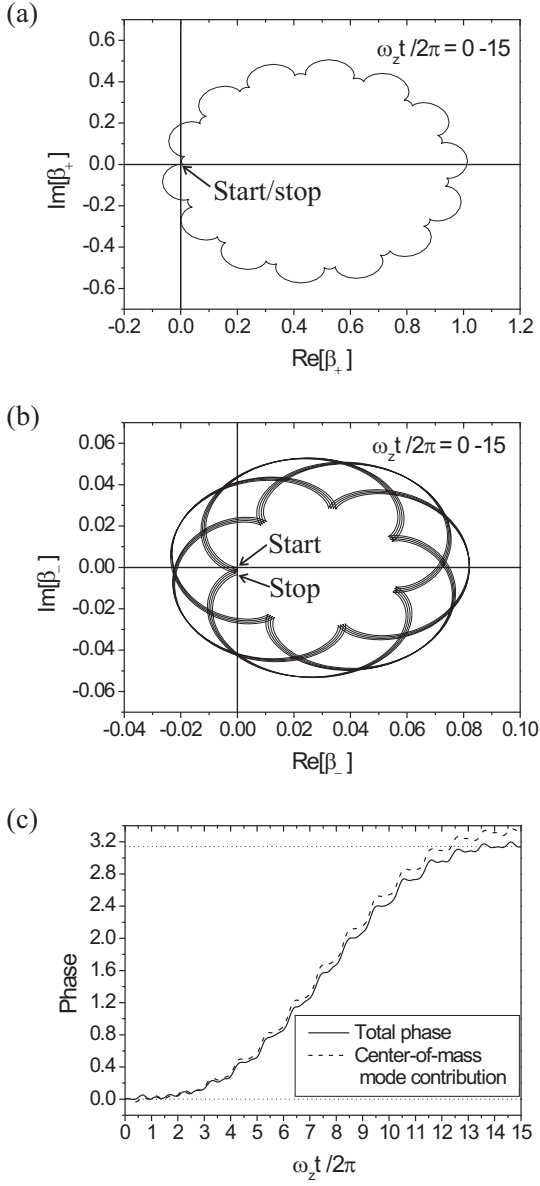}
\caption{Real and imaginary parts of the center-of-mass mode and
the breathing mode displacement $\beta_\pm(t)$ and the acquired
phase with $n=15$. The parameters are chosen such that an
effective phase shift equal to $\pi$ of the \upup\ state (see
Eq.~\eqref{eq6:pi-requirement} below) is obtained. (a) Parametric
plot showing $\left(\RE[\beta_+(t)],
\IM[\beta_+(t)]\right)=(p_+/\sqrt{\hbar
m\omz},-z_+/\sqrt{\hbar/m\omz})$ when the ions are in \downdown.
(b) Parametric plot showing $\left(\RE[\beta_-(t)],
\IM[\beta_-(t)]\right)=(p_-/\sqrt{\hbar
m\sqrt{3}\omz},-z_-/\sqrt{\hbar/m\sqrt{3}\omz})$ when the ions are
in \downup. In both (a) and (b) the phase-space trajectory starts
out from the origin and goes clockwise as time elapses. (c)
Acquired phase. The solid line is the total effective phase of
$2[\phi_+(\downarrow\downarrow,t)-\phi_-(\downarrow\uparrow,t)]$.
The dashed line is the center-of-mass mode contribution of
$2\phi_+(\downarrow\downarrow,t)$. The dotted lines indicate
phases of $0$ and $\pi$.}\label{fig6:gate-disp-phase}
\end{figure}

To describe the motion of the ions, we first rewrite
Eq.~\eqref{eq6:pot-energy} as follows
\begin{align}\label{eq6:two-ions-linearized}
  U&(z_{1, eq}, z_{2, eq}, \alpha_1, \alpha_2, t)=
  \frac{3}{4}m\omz^2\Delta z^2+\frac{1}{2}m\omz^2\left(z_+^2+3z_-^2\right)\\\nonumber &+U_{dip}({z_{1, eq}, \alpha_1}, t)+U_{dip}({z_{2, eq}, \alpha_2},t)\\\nonumber
  &-\frac{1}{\sqrt{2}}\left[F_{dip}(z_{1, eq}, \alpha_1, t)+F_{dip}(z_{2, eq}, \alpha_2,
  t)\right]z_+\\\nonumber &-\frac{1}{\sqrt{2}}\left[F_{dip}(z_{2, eq}, \alpha_2, t)-F_{dip}(z_{1, eq}, \alpha_1,
  t)\right]z_-,
\end{align}
where  $z_+=\frac{1}{\sqrt{2}}(z_2+z_1)\quad \text{and}\quad
z_-=\frac{1}{\sqrt{2}}(z_2-z_1-\Delta z)$ are the motional mode
coordinates for the so-called center-of-mass mode and the
breathing mode~\cite{James}, respectively, and $\Delta z
=z_{2,eq}-z_{1,eq}$ denotes the equilibrium distance between the
ions. Turning to a quantum mechanical description, we introduce
lowering (raising) operators $a$ ($a^\dagger$) and $b$
($b^\dagger$) for the mode coordinates, i.e.,
$\hat{z}_+=\sqrt{\hbar/(4m\omz)}(a+a^\dagger)$ and
$\hat{z}_-=\sqrt{\hbar/(4m\sqrt{3}\omz)}(b+b^\dagger)$, to find
the Hamiltonian
\begin{align}\label{eq:Hamiltonian}
H&=\hbar \omega_z a^{\dagger}a + \sqrt{3}\hbar \omega_z
b^{\dagger}b\\\nonumber & + U_{dip}(z_{1,eq},
\alpha_{1},t)+U_{dip}(z_{2,eq},\alpha_{2},t)\\\nonumber
&+f_{+}(\alpha_{1},\alpha_{2},t)(a+a^{\dagger})+
f_{-}(\alpha_{1},\alpha_{2},t)(b+b^{\dagger}),
\end{align}
where
\begin{align}\label{eq6:f+} f_{+}(\alpha_{1},\alpha_{2},t)=
-\sqrt{\frac{\hbar}{8m\omega_z}}[&F_{dip}(z_{1,eq},\alpha_{1},t)\\\nonumber
&+F_{dip}(z_{2,eq},\alpha_{2},t)] \intertext{and}\label{eq6:f-}
f_{-}(\alpha_{1},\alpha_{2},t)=-\sqrt{\frac{\hbar}{8m\sqrt{3}\omega_z}}[&F_{dip}(z_{2,eq},\alpha_{2},t)\\\nonumber
&-F_{dip}(z_{1,eq},\alpha_{1},t)]
\end{align}
are responsible for excitation of the center-of-mass mode and the
breathing mode, respectively. The time-evolution of the system can
be described by a unitary time-evolution operator $\mathcal{U}$,
which evolves state vectors $\Psi$ in time according to
$\Psi(t)=\mathcal{U}(t)\Psi(t=0)$. $\mathcal{U}(t)$ solves the
time dependent Schrödinger equation, $i\hbar
\dot{\mathcal{U}}(t)=H(t)\mathcal{U}(t),\ \mathcal{U}(0)=I$. Since
the Hamiltonian can be decomposed as a sum of commuting terms,
$\mathcal{U}$ can be expressed as a product
\begin{align}\label{eq6:wavefunction}
  \mathcal{U}=&\exp\left[-\frac{i}{\hbar}\int_0^t dt'\left[U_{dip}(z_{1,eq},
\alpha_{1},t')+U_{dip}(z_{2,eq},\alpha_{2},t')\right]\right]\\\nonumber&\times\mathcal{U}_+(t)\mathcal{U}_-(t),
\end{align}
where $\mathcal{U}_+(t)$ and $\mathcal{U}_-(t)$ are time-evolution
operators corresponding to the Hamiltonians
\begin{align} H_+=&\hbar \omega_z
a^{\dagger}a+f_{+}(\alpha_{1},\alpha_{2},t)(a+a^{\dagger})\\\intertext{and}H_-=&\sqrt{3}\hbar
\omega_z
b^{\dagger}b+f_{-}(\alpha_{1},\alpha_{2},t)(b+b^{\dagger}),
\end{align}
respectively. In the following we solve for $\mathcal{U}_+$ with
the understanding that the solution $\mathcal{U}_-$ can be
obtained from $\mathcal{U}_+$ simply by replacing $f_+$ with $f_-$
and \omz\ with $\sqrt{3}\omz$. To this end, we switch to the
interaction picture with respect to the Hamiltonian of the free
harmonic oscillator
\begin{align}
H_{int,+}=&e^{i\omz ta^\dagger
a}f_+(\alpha_{1},\alpha_{2},t)(a^\dagger+a)e^{-i\omz ta^\dagger
a}\\\nonumber =&f_{+}(\alpha_{1},\alpha_{2},t)(ae^{-i\omz
t}+a^{\dagger}e^{i\omz t})\\
\intertext{and make the Ansatz that} \mathcal{U}_{int,+}=&e^{i\omz
ta^\dagger
a}\mathcal{U}_+=e^{-\abs{\beta_+}^2/2}e^{i\phi_+}e^{i\beta_+^\ast
a^\dagger}e^{i\beta_+ a}.
\end{align}
From the Schrödinger equation for $\mathcal{U}_{int,+}$ it follows
that the harmonic oscillator phase-space displacement
$\beta_+(\alpha_1,\alpha_2,t)=p_+/\sqrt{m\hbar\omz}-iz_+/\sqrt{\hbar/(m\omz)}$,
with $p_+$ being the center-of-mass mode momentum, can be written
as
\begin{align}\label{eq6:beta-integral}
  \beta_+(\alpha_1,\alpha_2,t)=&-\frac{1}{\hbar}\int_{0}^{t}dt' f_+(\alpha_1,\alpha_2,t')e^{-i\omz t'}\\
  \intertext{and that the phase $\phi_+$ acquired due to
excitation by the force term $f_+$ is given by
}\label{eq6:phi-integral}
\phi_+(\alpha_1,\alpha_2,t)=&-\frac{1}{\hbar^2}\IM\Bigg[\int_{0}^{t}dt'
f_+(\alpha_1,\alpha_2,t')e^{-i\omz t'}\\\nonumber &\times
\left(\int_{0}^{t'}dt''f_+(\alpha_{1},\alpha_{2},t'')e^{i\omz
t''}\right)\Bigg].
\end{align}
The displacement of the ions is internal state dependent, which
leads to coupling (or entanglement) between the internal and the
motional states. For the gate operation, this is an undesired
effect and we shall therefore request the displacement to be zero
at the end of the gate operation, such that
$\mathcal{U}_{int,+}=e^{i\phi_+}$. In the implementation of the
Controlled-Z gate described below, we will take the
dipole-potential to be of the form
$U_{dip}(z_{i},\alpha_i,t)=U_{const}(z_{i})+U_{osc}(z_{i},\alpha_i)g(t)$,
$(i=1,2)$ in a time interval $[0,T]$ and zero otherwise,
specifically with $g(t)=0$ outside this interval. To avoid the
large internal state dependent phase factors in
Eq.~\eqref{eq6:wavefunction} we furthermore assume
$\int_0^Tg(t)dt=0$. The force term $f_+$ can be written as
$f_+(\alpha_1,\alpha_2,t)=f_{const}(z_{1,eq},z_{2,eq})+f_{osc}(z_{1,eq},z_{2,eq},\alpha_1,\alpha_2)g(t)$,
which together with Eq.~\eqref{eq6:beta-integral} implies that the
center-of-mass mode displacement at the end of the gate operation
can be written
\begin{equation}
\beta_+(T)=i\frac{f_{const}}{\hbar\omz}(1-e^{-i\omz
T})-i\frac{f_{osc}}{\hbar}\int_0^T dtg(t)e^{-i\omz t}.
\end{equation}
Both terms of this expression vanish if  $T$ is an integer number
$n$ of oscillation periods, $T=2\pi n/\omz$, and if the Fourier
transform
\begin{align}
\tilde{g}(\omz)&=\frac{1}{\sqrt{2\pi}}\int_{-\infty}^\infty
dtg(t)e^{-i\omz t}=\frac{1}{\sqrt{2\pi}}\int_0^T dtg(t)e^{-i\omz
t}\\\nonumber&=0.
\end{align}
From Eq.~\eqref{eq6:phi-integral} it can be shown that the phase
acquired during the gate operation can be expressed as
\begin{align}
\phi_+(T)=&\frac{1}{\hbar^2}\int
d\omega'\frac{\abs{\tilde{f_+}(\omega')}^2}{\omega'-\omz}\\\nonumber
=&C_1f_{const}^2+C_2f_{const}f_{osc}\tilde{g}(0)+\frac{1}{\hbar^2}\int
d\omega'\frac{\abs{\tilde{g}(\omega')}^2}{\omega'-\omz},
\end{align}
where $\tilde{f_+}(\omega)$ is the Fourier transform of $f_+(t)$
and $C_1$ and $C_2$ are constants. The first term on the r.h.s. is
irrelevant, since it does not depend on the internal state of the
ions and the second term vanishes since we have required
$\tilde{g}(0)=(2\pi)^{-1/2}\int_0^Tg(t)dt=0$. The interesting term
is the last one, from which we observe that the closer the
characteristic frequencies of the function $g(t)$ are to the
oscillation frequency \omz\ (or $\sqrt{3}\omz$ for the breathing
mode), the larger is the accumulated phase. It is thus a natural
choice to adopt a harmonic time dependent force that oscillates
with just one oscillation cycle less (or more) than the trapping
motion during the interaction time $T$. In
Fig.~\ref{fig6:gate-disp-phase}(a), one example of a phase-space
trajectory of the center-of-mass mode of two ions in the
$\downdown$ state is shown. The duration of the interaction is
taken to be 15 trapping periods, and 14 periods of the applied
periodic force. We observe that the net displacement vanishes at
the end of the gate operation. For the same interaction,
Fig.~\ref{fig6:gate-disp-phase}(b) shows the phase-space
trajectory of the breathing mode with the ions being in the
\down\up\ state. The breathing mode frequency of $\sqrt{3}\omz$ is
far-off resonant with the frequency of the applied force and hence
the breathing mode is much less excited than the center-of-mass
mode. Note that since $15\sqrt{3}=25.98\sim 26$, this oscillator
mode undergoes an almost integer number of oscillations and hence
it experiences a nearly vanishing net displacement at the end of
the gate operation. Other good choices for the duration of the
interaction in units of the trapping period are 56
($56\sqrt{3}=96.995\sim 97$), and 209 ($209\sqrt{3}=361.9986 \sim
362$). The phases $\phi_{\pm}$ are equal to the areas of the
$(z_{\pm},p_{\pm})$ phase-space trajectories in units of $\hbar$,
and they depend on the internal state of both ions, as needed for
a two-qubit gate. Fig.~\ref{fig6:gate-disp-phase}(c) shows the
build-up of the phase during the interaction. The parameters have
been chosen to ensure an effective phase shift of $\pi$ on the
\upup\ state (see Eq.~\eqref{eq6:pi-requirement} below) as desired
for the Controlled-Z gate. In the following section on the
physical implementation of the scheme, analytical expressions for
the acquired phases are provided.

\section{Implementation in alkaline earth ions}\label{sec:implementation}
\subsection{Atom-light coupling}
\begin{figure}
  \centering
  \includegraphics[width=0.3\textwidth]{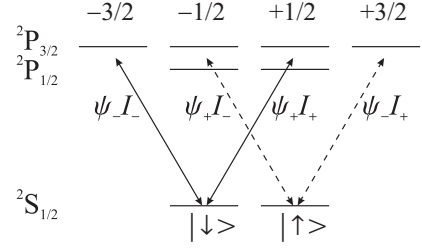}
  \caption{Relevant energy-levels and transitions in alkaline earth ions
  (e.g.,\cafp, \srp, and \bap) for calculating the dipole potential for the
  qubit states $\down=\state{2}{S}{1/2}(-1/2)$ (solid lines) and
  $\up=\state{2}{S}{1/2}(+1/2)$ (dashed lines) due to a dipole force inducing laser beam containing $\sigma^+$-- and
  $\sigma^-$--polarized components with intensities $I_+$ and $I_-$, respectively.
  $\psi_+$ and $\psi_-$ are defined in the text.}\label{fig6:implementation}
\end{figure}

Since alkaline earth ions are among the most prominent atomic ions
for doing quantum logic, we will now focus on experimental
realizations of the Controlled-Z gate based on such ions with
presentations of specific results for \cafp\ and \bap. We choose
the qubit states to be $\down=n\state{2}{S}{1/2}(-1/2)$ and
$\up=n\state{2}{S}{1/2}(+1/2)$, on which single qubit operations
can be performed using Raman transitions via the $
n\state{2}{P}{1/2}$ level. Assuming that the frequency \omsub{L}
of the dipole force inducing laser is close to or below the
transition frequencies $\omsub{1/2}$ and $\omsub{3/2}$ of the
$n\state{2}{S}{1/2}$--$n\state{2}{P}{1/2}$ and the
$n\state{2}{S}{1/2}$--$n\state{2}{P}{3/2}$ transitions,
respectively, we may to a good approximation only consider
contributions from these two transitions to the dipole potential
(as long as \omsub{L} is not in the immediate vicinity of the
transition frequencies of the weak
$n\state{2}{S}{1/2}-(n-1)\state{2}{D}{3/2,5/2}$ electric
quadrupole transitions). Expanding the dipole force inducing beam
into $\sigma^+$- and $\sigma^-$-polarized light components with
respect to its propagation axis (the quantization axis), the
contributions to the dipole potential for the two states \down\
and \up\ are those associated with the transitions indicated in
Fig.~\ref{fig6:implementation}. The respective dipole potentials
can consequently be written
\begin{equation}\label{eq6:dipole-pot's}
U_\downarrow=\psi_+I_++\psi_-I_-\quad\text{and}\quad
U_\uparrow=\psi_-I_++\psi_+I_-,
\end{equation}
where $I_\pm$ is the intensity of the $\sigma^+$- and
$\sigma^-$-polarized components, respectively, and where
\begin{align} \label{eq6:psi's}
\psi_{+}=&\frac{3\pi
c^{2}}{2}\Bigg[\frac{2\Gamma_{1/2}}{3\omega_{1/2}^{3}}\left(\frac{1}{\omega_{1/2}-\omega_{L}}+\frac{1}{\omega_{1/2}+\omega_{L}}\right)\\\nonumber &+\frac{\Gamma_{3/2}}{3\omega_{3/2}^{3}}\left(\frac{1}{\omega_{3/2}-\omega_{L}}+\frac{1}{\omega_{3/2}+\omega_{L}}\right)\Bigg]\\
\intertext{and} \psi_{-}=&\frac{3\pi
c^{2}}{2}\frac{\Gamma_{3/2}}{\omega_{3/2}^{3}}\left(\frac{1}{\omega_{3/2}-\omega_{L}}+\frac{1}{\omega_{3/2}+\omega_{L}}\right)
\end{align}
depend only on the properties of the ion and the frequency
\omsub{L} of the dipole force inducing laser. Here, $\Gamma_{1/2}$
and $\Gamma_{3/2}$ are the transition strengths of the
$n\state{2}{S}{1/2}$--$n\state{2}{P}{1/2}$ and the
$n\state{2}{S}{1/2}$--$n\state{2}{P}{3/2}$ transitions,
respectively. As can immediately be seen, the force derived from
this dipole potential will be different for the two qubit states,
whenever the intensities of the two polarization components
differ. Hence, by introducing a sinusoidal temporal variation of
the intensity of the polarization components given by
\begin{equation}\label{eq6:pol-intensities}
  I_\pm(z,t)=\frac{1}{2}I(z)[1\pm\sin(\Omega t)],
\end{equation}
which can be done by, e.g., using an electro optic phase
modulator, the situation considered in Sec.~\ref{sec:theory} can
be obtained.

From Sec.~\ref{sec:theory} we thus identify $g(t)=\sin(\Omega t)$
for $0\leq t\leq T$. As for the choice of $\Omega$, we recall the
requirements $T=2\pi n/\omz$ with $n$ integer,
$\tilde{g}(0)=\tilde{g}(\omz)=\tilde{g}(\sqrt{3}\omz)=0$ and that
the characteristic frequency of $g(t)$, i.e., $\Omega$, should be
close to $\omz$. Consequently, a good choice is
$\Omega=(1-1/n)\omz$ with $n\gg 1$, as in the example of
Fig.~\ref{fig6:gate-disp-phase} where $n=15$. The Fourier
transform of $g(t)$ on the interval $[0,T]$ contains two terms
which are proportional to
$\sin[(\omega-\Omega)T/2]/(\omega-\Omega)$ and
$\sin[(\omega+\Omega)T/2]/(\omega+\Omega)$, respectively. These
sinc-functions peak at $\omega=\Omega$ and $\omega=-\Omega$, while
they have exact zeros at $\omega=0$ and $\omega=\omz$.
Furthermore, they are suppressed at $\omega=\sqrt{3}\omz$,
especially when $\sqrt{3}n$ is close to an integer.

\subsection{Phase shifts}
In order to calculate the displacement and the acquired phase for
the center-of-mass mode and the breathing mode, respectively, for
the specific intensity variation in
Eq.~\eqref{eq6:pol-intensities}, the $f$-functions entering in the
integrals in Eqs.~\eqref{eq6:beta-integral} and
~\eqref{eq6:phi-integral} have to be determined using the
definitions in Eq.~\eqref{eq6:f+} or \eqref{eq6:f-}. For $f_+$,
which is responsible for excitation of the center-of-mass mode, we
find
\begin{align}\label{eq:f+'s}
f_{+}(\downarrow\downarrow)=&-\left[f_{0+}+f_{1+}\sin(\Omega t)\right],\\
f_{+}(\downarrow\uparrow)=&-\left[f_{0+}+f_{2+}\sin(\Omega t)\right],\\
f_{+}(\uparrow\downarrow)=&-\left[f_{0+}-f_{2+}\sin(\Omega t)\right],\\
f_{+}(\uparrow\uparrow)=&-\left[f_{0+}-f_{1+}\sin(\Omega t)\right],\\
\intertext{where}
f_{0+}=&\sqrt{\frac{\hbar}{8m\omz}}(\tilde{F_{1}}+\tilde{F_{2}})(\psi_{+}+\psi_{-}),\\\label{eq6:f1+}
f_{1+}=&\sqrt{\frac{\hbar}{8m\omz}}(\tilde{F_{1}}+\tilde{F_{2}})(\psi_{+}-\psi_{-}),\\\label{eq6:f2+}
f_{2+}=&\sqrt{\frac{\hbar}{8m\omz}}(\tilde{F_{1}}-\tilde{F_{2}})(\psi_{+}-\psi_{-})\\
\intertext{and}\label{eq6:F-tilde}
\tilde{F_i}=&-\frac{1}{2}\frac{\partial I(z)}{\partial
z}\bigg\vert_{z=z_{i,eq}}\qquad(i=1,2).
\end{align}
The full time-dependent expressions $\beta(t)$ and $\phi(t)$,
which were used for obtaining the plots in
Fig.~\ref{fig6:gate-disp-phase}, can be found by carrying out the
integrals in Eqs.~\eqref{eq6:beta-integral}
and~\eqref{eq6:phi-integral} for the center-of-mass mode as well
as the breathing mode~\cite{staanum-thesis}. Here we only state
the acquired phases at the end of the gate operation, which for
the center-of-mass mode are given by
\begin{align}\label{eq6:phi-T}
  \phi_+(\downarrow\downarrow,T)=&\phi_+(\uparrow\uparrow,T)=\frac{f_{1+}^2}{(2\hbar)^2}\frac{2\omz
  T}{\omz^2-\Omega^2}\approx\frac{f_{1+}^2}{(\hbar\omz)^2}\frac{n^2\pi}{2},\\
   \phi_+(\downarrow\uparrow,T)=&\phi_+(\uparrow\downarrow,T)=\frac{f_{2+}^2}{(2\hbar)^2}\frac{2\omz
  T}{\omz^2-\Omega^2}\approx\frac{f_{2+}^2}{(\hbar\omz)^2}\frac{n^2\pi}{2}
\end{align}
neglecting a term which is independent of the internal state and
assuming $n\gg 1$. These phases scale quadratically with $n$ with
one factor of $n$ originating from the time $T$ and the other
originating from the denominator, due to the fact that $\Omega$
was chosen to be near-resonant with \omz. The phases acquired due
to excitation of the breathing mode are likewise found to be
\begin{align}\label{eq6:phi-T-breathupup}
  \phi_-(\downarrow\downarrow,T)=&\phi_-(\uparrow\uparrow,T)=\frac{f_{1-}^2}{(2\hbar)^2}\frac{2\sqrt{3}\omz
  T}{3\omz^2-\Omega^2}\approx\frac{f_{1-}^2}{(\hbar\omz)^2}\frac{\sqrt{3}n\pi}{2}\\\label{eq6:phi-T-breathupdown}
   \phi_-(\downarrow\uparrow,T)=&\phi_-(\uparrow\downarrow,T)=\frac{f_{2-}^2}{(2\hbar)^2}\frac{2\sqrt{3}\omz
  T}{3\omz^2-\Omega^2}\approx\frac{f_{2-}^2}{(\hbar\omz)^2}\frac{\sqrt{3}n\pi}{2}\end{align}
where
\begin{equation}\label{eq6:f1-2-}
   f_{1-}=f_{1+}\frac{\tilde{F_{2}}-\tilde{F_{1}}}{\sqrt[4]{3}(\tilde{F_{2}}+\tilde{F_{1}})}\quad\text{and}\quad
   f_{2-}=f_{2+}\frac{\tilde{F_{2}}+\tilde{F_{1}}}{\sqrt[4]{3}(\tilde{F_{2}}-\tilde{F_{1}})}.
\end{equation}
These phases only scale linearly with $n$, since the rotation
frequency $\Omega$ is off-resonant with the breathing mode
frequency of $\sqrt{3}\omz$. In Eqs.~\eqref{eq6:phi-T-breathupup}
and~\eqref{eq6:phi-T-breathupdown}, terms which are smaller than
the stated terms by a factor of $n$ and further suppressed if
$\sqrt{3}n$ is close to an integer have been neglected.

The combined effect of the above phase shifts is equivalent to a
single phase shift
$\phi_\pm={\phi_\pm(\downarrow\downarrow)}-{\phi_\pm(\downarrow\uparrow)}-{\phi_\pm(\uparrow\downarrow)}+{\phi_\pm(\uparrow\uparrow)}=2[{\phi_\pm(\downarrow\downarrow)}-{\phi_\pm(\downarrow\uparrow)}]$
of the \upup\ state~\cite{Sasura-Steane}. Thus, to make a
Controlled-Z gate, we require that
\begin{align}\label{eq6:pi-requirement}
\pi=&2\left[\phi_+(\downarrow\downarrow,T)+\phi_-(\downarrow\downarrow,T)-\phi_+(\downarrow\uparrow,T)-\phi_-(\downarrow\uparrow,T)\right]\\\nonumber
\approx&\frac{\pi
n^2}{(\hbar\omz)^2}\left[f_{1+}^2-f_{2+}^2+\frac{\sqrt{3}}{n}\left(f_{1-}^2-f_{2-}^2\right)\right].
\end{align}
\section{Realization of the Controlled-Z gate using gaussian
laser beams}\label{sec:Gaussian beams} \subsection{Intensity
requirements and off-resonant scattering rates}

\begin{figure}[!htbp]
\centering\includegraphics[width=\linewidth]{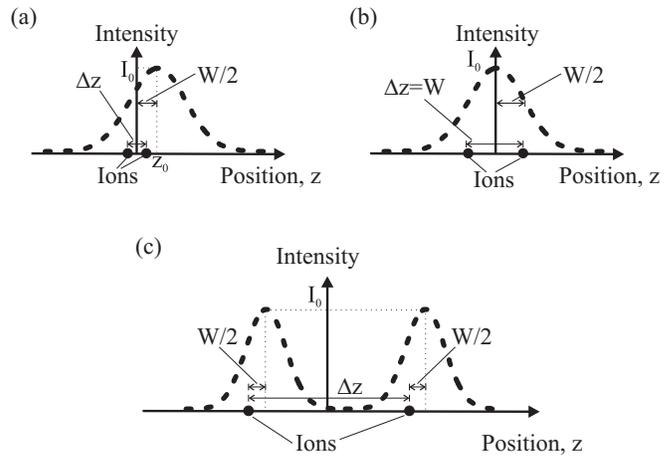} 
\caption{Three possible configurations for realizing the two-ion
gate considered in the text. (a) Two ions positioned at distances
of $W/2\pm\Delta z/2$ from the center of a Gaussian laser beam.
(b) Two ions positioned at distances of $\pm W/2=\pm\Delta z/2$
from the center of a Gaussian laser beam. (c) Two ions, each
positioned at a distance of $W/2$ from the center of a tightly
focussed Gaussian laser beam.}\label{fig6:beams-ions}
\end{figure}

In the following we will consider possible realizations of a
Controlled-Z gate in a two ion string using one or two dipole
force inducing laser beams propagating perpendicular to the trap
axis as sketched in Fig.~\ref{fig6:beams-ions}. In the first case
shown in Fig.~\ref{fig6:beams-ions}(a), we assume that the
equilibrium distance $\Delta z$ between the two ions is smaller
than the waist of the laser beam, such that they feel essentially
the same strong dipole force. The situation in
Fig.~\ref{fig6:beams-ions}(b) corresponds to a case where the size
of the force on the two ions are the same but of opposite sign for
identical internal states. Finally, Fig.~\ref{fig6:beams-ions}(c)
shows a realization similar to the first case, but with
application of two laser beams.

Since in all cases we will assume the ions to be situated close to
one of the points in the intensity profile where the induced
dipole force is largest, much of the analysis is similar for the
three situations. Hence, in the following we will focus on the
situation in Fig.~\ref{fig6:beams-ions}(a) in order to establish
formal equations. Assuming the ion string is centered at $z=0$, we
may write the intensity of the dipole force inducing beam as

\begin{equation}\label{eq6:Gaussian-profile}
I(z)=I_0e^{-2\left(z-z_0\right)^2/W^2},
\end{equation}
where $z_0=W/2$ is the center of the dipole force inducing beam,
$W$ is the beam waist and $I_0$ is the peak intensity. For the
gate operation two highly relevant and linked parameters are the
required peak intensity $I_0$ and the fidelity loss due to
off-resonant scattering events. The required peak intensity enters
in Eq.~\eqref{eq6:pi-requirement} through the $f$-functions. For a
given wavelength and a given choice of beam waist the required
peak intensity can be determined. Using the definitions in
Eqs.~\eqref{eq6:f1+},~\eqref{eq6:f2+},~\eqref{eq6:F-tilde}
and~\eqref{eq6:f1-2-}, we find
$\tilde{F_1}\approx\tilde{F_2}\approx e^{-1/2}I_0/W$,
$f_{2+}\approx f_{1-}\approx 0$ and a pair of non-zero expressions
for $f_{1+}$ and $f_{2-}$, which together with
Eq.~\eqref{eq6:pi-requirement} yields the required peak intensity
\begin{equation}\label{eq6:required-intensity}
  I_0\approx\sqrt{\frac{2e^1\hbar\omz^3
  m W^2}{n^2(\psi_+-\psi_-)^2}},
\end{equation}
where only the leading term in $n$ has been retained. The scaling
with the various parameters is intuitively reasonable. First, it
has already been discussed that large $n$ leads to a large phase
pick-up and hence a low intensity requirement. Second, the larger
the difference $\psi_+-\psi_-$, the larger is the dipole force
difference for \down\ and \up. Third, if the waist is small, the
dipole force is large, which in turn reduces the required
intensity. Note also that the required laser \emph{power} ($\sim
I_0W^2$) is proportional to $W^3$, which makes a small waist very
attractive.

Knowing the required intensity, we can now determine the average
probability $P_{sc}$ for a scattering event from one of the two
ions during the gate operation. Formally, we can write
$P_{sc}\approx \Gamma_{sc}T$ ($\Gamma_{sc}T\ll 1$), where
$\Gamma_{sc}$ is the average total scattering rate of the two
ions. Assuming an equal average population of the two internal
states and an intensity of $e^{-1/2}I_0$ at the position of the
ions, the dominant scattering from the $n\state{2}{P}{1/2}$- and
the $n\state{2}{P}{3/2}$-state yields
\begin{align}\label{eq6:infidelity}
  P_{sc}\approx&\frac{\tilde{\Gamma}_{sc}}{\psi_+-\psi_-}\sqrt{8\pi^2\hbar\omz
  mW^2}\\
  \intertext{where}\label{eq6:gamma-tilde}
  \tilde{\Gamma}_{sc}=&\frac{3\pi c^2\omsub{L}^3}{2\hbar}\Bigg[\frac{\Gamma_{1/2}^2}{\omsub{1/2}^6}\left(\frac{1}{\omsub{1/2}-\omsub{L}}+\frac{1}{\omsub{1/2}+\omsub{L}}\right)^2\\\nonumber &+\frac{\Gamma_{3/2}^2}{\omsub{3/2}^6}\left(\frac{1}{\omsub{3/2}-\omsub{L}}+\frac{1}{\omsub{3/2}+\omsub{L}}\right)^2\Bigg].
\end{align}
The dependency of $P_{sc}$ on the internal structure of the ion
and the wavelength of the dipole force inducing laser is contained
in the front factor of $\tilde{\Gamma}_{sc}/(\psi_+-\psi_-)$,
showing that $P_{sc}$ can be minimized either by making
$\tilde{\Gamma}_{sc}$ small or $\psi_+-\psi_-$ large.
$\tilde{\Gamma}_{sc}$ becomes small in the limit
$\omsub{L}\ll\omsub{1/2},\omsub{3/2}$ due to the factor of
$\omsub{L}^3$, however, in the same limit $\psi_+-\psi_-$ is
proportional to the \emph{difference}
$\Gamma_{1/2}\omsub{1/2}^{-4}-\Gamma_{3/2}\omsub{3/2}^{-4}$, which
is also small. Alternatively, if
$\omsub{1/2}<\omsub{L}<\omsub{3/2}$, $\psi_+-\psi_-$ can become a
\emph{sum} of two positive terms, however, for
$\tilde{\Gamma}_{sc}$ to be small in this case, a large
fine-structure splitting is required. Both in the far-off resonant
case and when the dipole force inducing laser is tuned in between
the fine-structure levels, the \bap\ ion turns out to be more
attractive than the \cafp\ ion. Another candidate ion is \srp\
which for the present gate proposal is less attractive than \bap\
but more attractive than \cafp, however, in the following we only
consider \cafp\ and \bap.

\subsection{Experimental parameters for \cafp\ and \bap}
To get more quantitative numbers out, we now consider the cases of
two \cafp\ or two \bap\ ions in a trap with $\omz=2\pi\times
1\unit{MHz}$, which leads to an equilibrium distance $\Delta z$ of
5.6\unit{\mu m} and 3.7\unit{\mu m} between two ions,
respectively. First we consider the situation depicted in
Fig.~\ref{fig6:beams-ions}(a) for $\Delta z\ll W$, such that the
dipole forces on the two ions are equal. To fulfill this we choose
a waist of $30\unit{\mu m}$. Finally, by choosing $n=15$
($\sqrt{3}n\approx 26$) the intensity and the scattering rate can
be calculated for a given wavelength of the dipole force inducing
laser beam. For \cafp\ and \bap\ with the dipole force inducing
laser tuned either in between the fine-structure levels or far red
detuned, we find the following values for (Power, $P_{sc}$,
$\lambda_L$), where $\lambda_L$ is the dipole force inducing laser
wavelength. \cafp: (8\unit{W}, 30$\%$, 395.1\unit{nm}) and
($\sim$0.5\unit{MW}, $<4\%$, $>$1500\unit{nm}). \bap: (86\unit{W},
$6\%$, 474.5\unit{nm}) and ($\sim$33\unit{kW}, $<1.2\%$,
$>$1000\unit{nm}). Clearly, this is not very promising for
experimental realizations. However, since the main problem is the
requirement that $W\gg\Delta z$, the situation depicted in
Fig.~\ref{fig6:beams-ions}(b) is more favorable. Here $W=\Delta z$
with the ions positioned symmetrically around the center of the
dipole force inducing beam. Since we do not have the requirement
that $\Delta z\ll W$ we can choose a lower trap frequency of
$\omz=2\pi\times 200\unit{kHz}$, which implies that $\Delta z$ is
equal to 16.4\unit{\mu m} for \cafp\ and 10.9\unit{\mu m} for
\bap\ and which yields the following values for (Power, $P_{sc}$,
$\lambda_L$). \cafp: (120\unit{mW}, 8\%, 395.1\unit{nm}) and
($\sim$6.5\unit{kW}, $<0.9\%$, $>$1500\unit{nm}). \bap:
(360\unit{mW}, 1\%, 474.5\unit{nm}) and ($\sim$140\unit{W},
$<0.2\%$, $>$1000\unit{nm}). For high fidelity gates the required
power is experimentally still too demanding.

\begin{figure*}[htbp]
\centering
\makebox[0.42\linewidth][l]{(a)}
\makebox[0.42\linewidth][l]{(b)}\vspace{-6mm}\\
\includegraphics[width=0.42\linewidth]{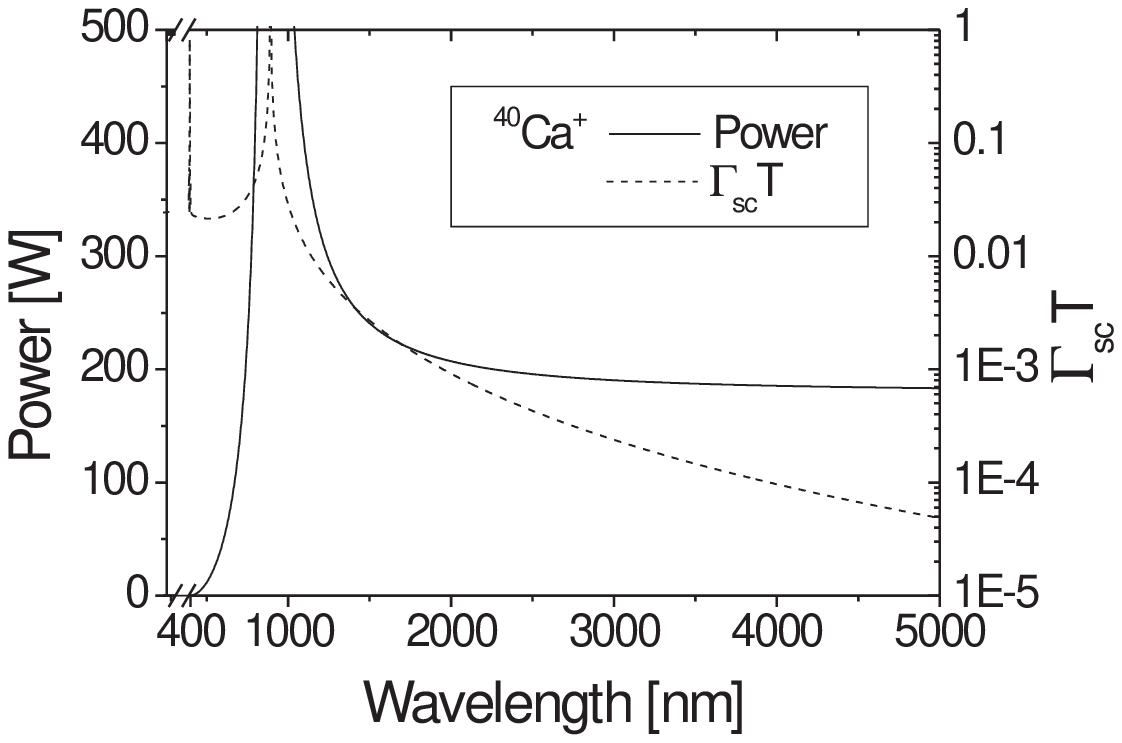}
\includegraphics[width=0.42\linewidth]{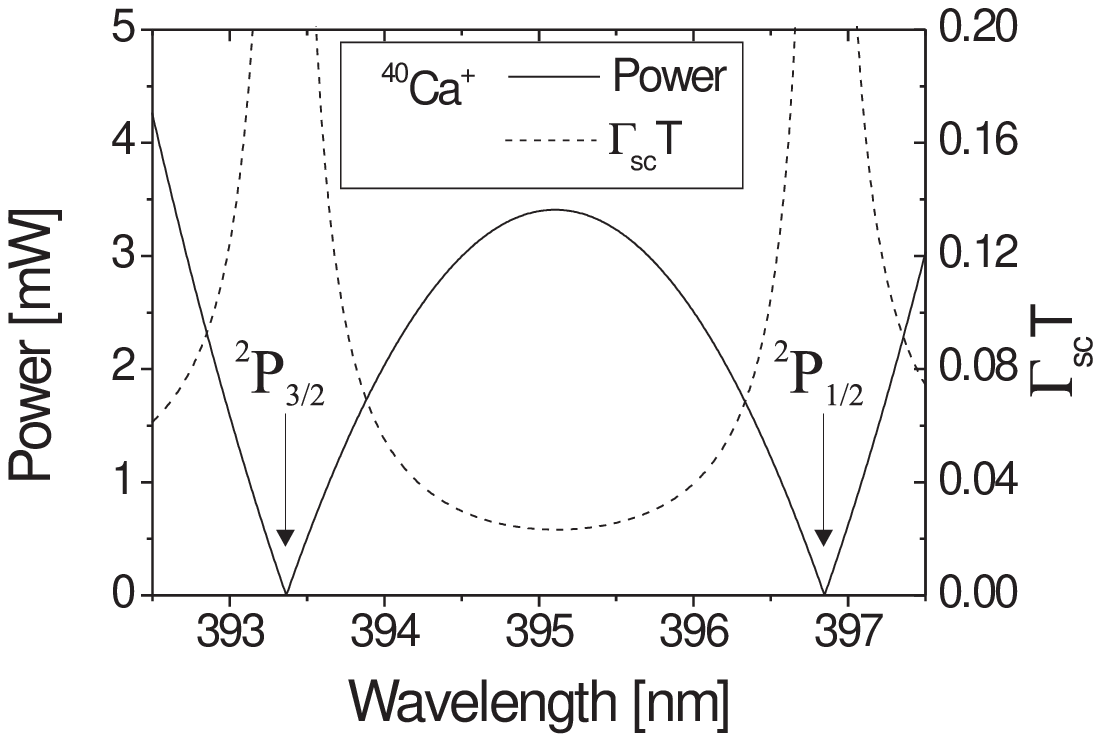}\\\vspace{-3mm}
\makebox[0.42\linewidth][l]{(c)}
\makebox[0.42\linewidth][l]{(d)}\vspace{-6mm}\\
\includegraphics[width=0.42\linewidth]{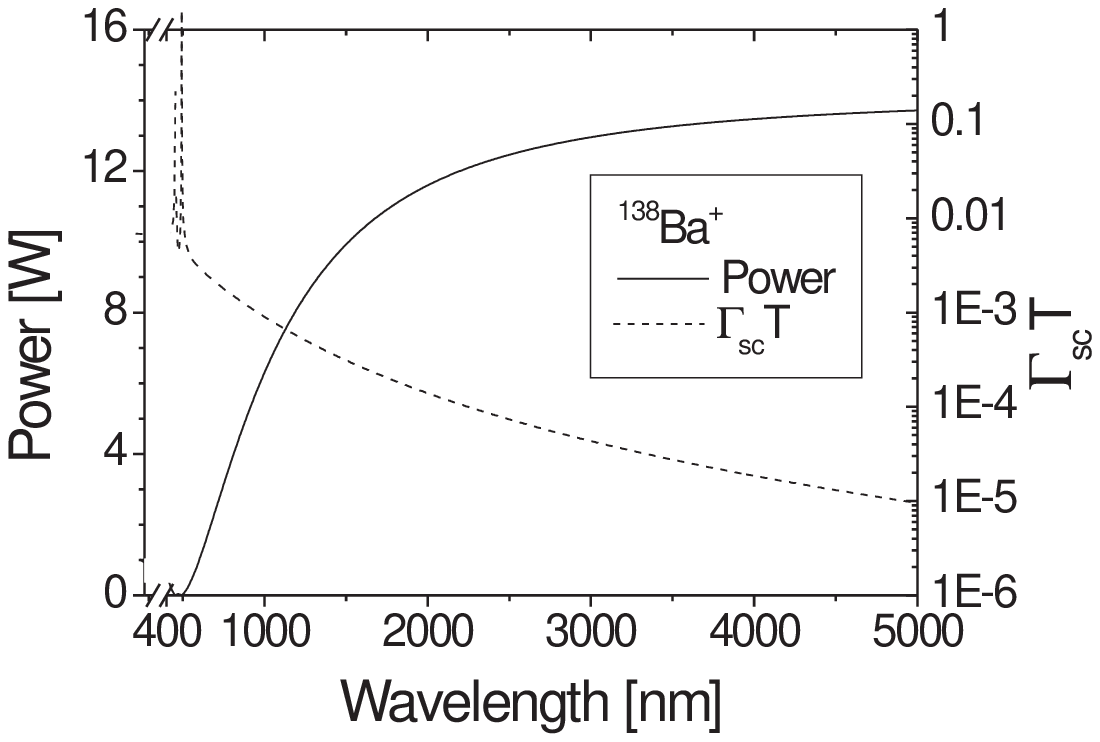}
\includegraphics[width=0.42\linewidth]{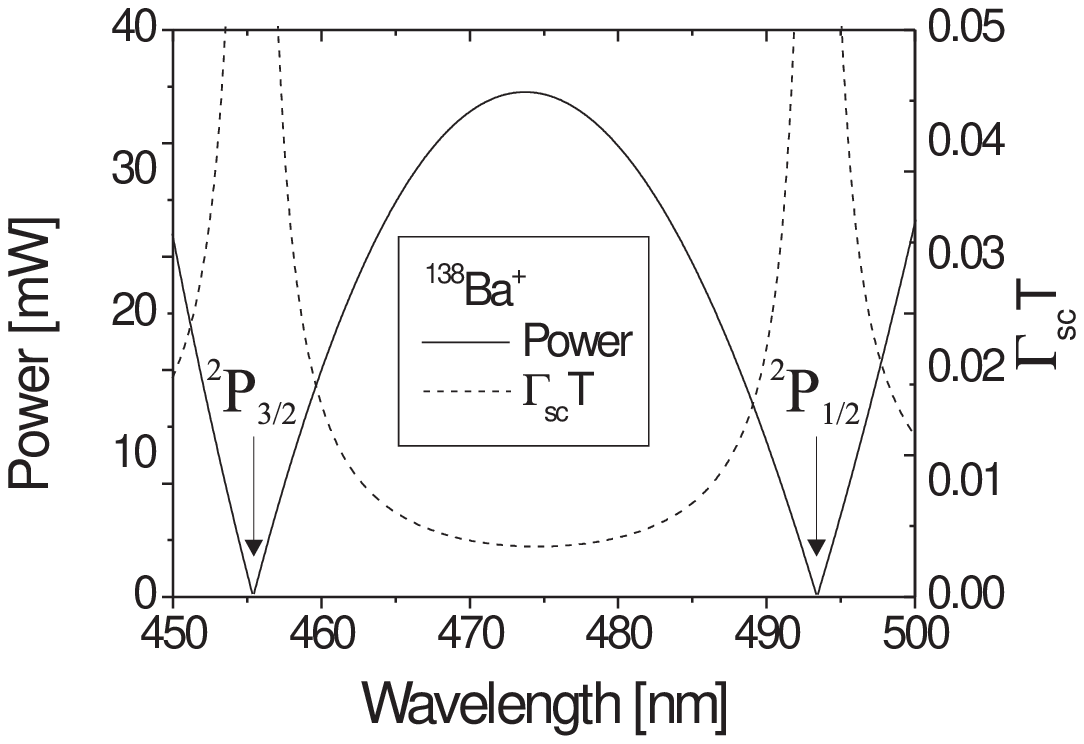}\\\vspace{-5mm}
\caption{Required power and $\Gamma_{sc}T$ vs. wavelength of the
dipole force inducing laser for \cafp\ and \bap. In all plots
$\omz=2\pi\times 200\unit{kHz}$, $W=5\unit{\mu m}$, $n=15$ and
$T=75\unit{\mu s}$. (a) \cafp, far red detuned laser. The
divergences near 900\unit{nm} are due to a cancellation of
$\psi_+-\psi_-$. (b) \cafp, laser wavelength in the vicinity of
the $4\state{2}{S}{1/2}$--$4\state{2}{P}{1/2}$ and
$4\state{2}{S}{1/2}$--$4\state{2}{P}{3/2}$ transition wavelengths,
which are indicated by the symbols \state{2}{P}{1/2} and
\state{2}{P}{3/2}. (c) \bap, far red detuned laser. (d) \bap,
laser wavelength in the vicinity of the
$6\state{2}{S}{1/2}$--$6\state{2}{P}{1/2}$ and
$6\state{2}{S}{1/2}$--$6\state{2}{P}{3/2}$ transition
wavelengths.}\label{fig6:gate-plots}
\end{figure*}

Finally, we consider the situation presented in
Fig.~\ref{fig6:beams-ions}(c) where the waist of the applied laser
beams is not directly related to the equilibrium distance between
the ions and hence rather tightly focussed beams can be applied.
As long as the beam waists are larger than the thermal excursion
of the ions from their equilibrium positions, the equations
derived above for the case of Fig.~\ref{fig6:beams-ions}(a) is
immediately applicable. For $\omz=2\pi\times 200\unit{kHz}$ and
$W=5\unit{\mu m}$, we plot in Fig.~\ref{fig6:gate-plots} the
required laser power and $\Gamma_{sc}T\approx P_{sc}$ for \cafp\
and \bap\ as a function of the dipole force inducing laser
wavelength. The plots in Fig.~\ref{fig6:gate-plots}(a,c) extend to
a wavelength of $5\unit{\mu m}$ in order to show the long
wavelength behavior, however, one should keep in mind that due to
diffraction it is technically demanding to obtain a 5\unit{\mu m}
waist for the longest wavelengths. In all plots $n=15$ was chosen,
which gives a very reasonable gate time of $T=75\unit{\mu s}$.

For \cafp, we see from Fig.~\ref{fig6:gate-plots}(a) that a high
power of $\sim200\unit{W}$ is required in the long wavelength
limit for the considered parameters. However, by choosing $n=209$
instead of 15, the required power drops to $\sim15\unit{W}$ at the
cost of an increased gate time of $T=1\unit{ms}$. Considering the
challenge of diffraction a laser at a wavelength of 2\unit{\mu m}
would be favorable in the long wavelength limit. Fortunately, the
required power of $\sim 15\unit{W}$ is easily achievable with
commercial Thulium fiber lasers operating in the range of 1.75 -
2.2\unit{\mu m} \endnote{High power single mode Thulium fiber
laser from IPG photonics (TLR-series),
\texttt{www.ipgphotonics.com}.}. Hence, high fidelity gates with
$P_{sc}\approx10^{-3}$ should be possible. Other interesting
lasers in the far-off resonant regime are the Nd:YAG (yttrium
aluminium garnet) laser at 1064\unit{nm} and the $\text{CO}$-laser
near 5\unit{\mu m}, but the relatively high scattering probability
and diffraction limitations, respectively, makes these lasers less
attractive than the Thulium fiber laser. A wavelength near
395\unit{nm} is furthermore attractive for \cafp\ (see
Fig.~\ref{fig6:gate-plots}(b)), since the needed power of only
$\sim3\unit{mW}$ easily can be obtained. The scattering
probability is a few percent, which is acceptable for a first
demonstration, but not good enough for implementation of error
correcting schemes. By a further reduction of the beam waist,
e.g., by placing a lens system inside the vacuum chamber where the
ion trap is situated, it should be possible to focus to below
1\unit{\mu m} ~\cite{Schlosser-microdipoletrap} and hence reduce
the scattering rate by a factor of $5-10$.

Due to the larger fine-structure splitting of \bap, the scattering
probability is already below 1\% for a wavelength around
475\unit{nm} (see Fig.~\ref{fig6:gate-plots}(d)) between the
$6\state{2}{S}{1/2}$--$6\state{2}{P}{1/2},\state{2}{P}{3/2}$
transition wavengths. The low laser power needed in this
wavelength region can easily be supplied by frequency doubled
diode laser systems. Note also that light from an Argon-ion laser
at a wavelength of 488\unit{nm} could be a reasonable possibility.
However, as for \cafp\, a Thulium fiber laser seems to be most
ideal, since high fidelity gates with $P_{sc}\approx10^{-4}$ (a
value comparable to the threshold value required for
fault-tolerant quantum
computation~\cite{Steane-errorcorrectionII}) should be feasible
with laser powers of about 14\unit{W} for the case with $n=15$ and
$T=75\unit{\mu s}$ as shown in Fig.~\ref{fig6:gate-plots}(c). In
the long wavelength limit, the 1064\unit{nm} wavelength of a
Nd:YAG laser could also be a possibility since a scattering
probability of $\sim10^{-3}$ at a laser power of $\sim7\unit{W}$
is achievable. It should be noted that in all cases discussed
above, the required power, the scattering probability and the gate
time can be adjusted by changing $n$, \omz\ and $W$.
%
\subsection{Error sources}\label{sec:errors} When we considered
the realization of the gate schemes above, we already discussed
the effect of spontaneous emission on the fidelity of the gate
operations. There are, however, other experimental error sources
which will also influence the final fidelity of an actual
realization. A key element in the present gate proposal is the
polarization rotation, which serves to remove phases due to
internal state dependent Stark shifts. Since the Stark shifts are
of first order in intensity and the phase acquired by the change
in Coulomb energy only originates from second order effects, even
small errors in the laser parameters may be very critical to the
actual gate operation. In the following subsections, errors due to
non-perfect balancing of the two polarization components of the
dipole force inducing laser beam, timing errors as well as power,
position and frequency-fluctuations of the dipole force inducing
laser are discussed.
\subsubsection{Polarization errors} In case there is
an imbalance of the intensity in the two polarization components,
such that
\begin{equation}\label{eq6:bad-polarization}
I_\pm(z,t)=\frac{1}{2}I(z)[1\pm\epsilon_p][1\pm\sin(\Omega t )],
\end{equation}
where $\epsilon_p$ accounts for the imbalance, there will be two
extra terms in the dipole potential. One term
[$\propto\epsilon_p\sin(\Omega t)$] enters with the same sign in
$U_\downarrow$ and $U_\uparrow$ and hence it does not give rise to
any Stark-shift induced phase-difference between \down\ and \up.
The other term, $\epsilon_p I(z)(\psi_+-\psi_-)/2$, enters with a
different sign in $U_\downarrow$ and $U_\uparrow$, which leads to
a phase-difference of $\Delta\phi=\epsilon_p
I(z)(\psi_+-\psi_-)T/\hbar$ between \down\ and \up\ in each ion at
the end of the gate operation. Since this difference should be
much smaller than the desired phase-shift of $\pi$, we find by
using the expression for $\Delta\phi$,
Eqs.~\eqref{eq6:Gaussian-profile},~\eqref{eq6:required-intensity}
and $T=2\pi n/\omz$ the condition
\begin{equation}\label{eq6:bad-polarization condition}
  \epsilon_p\ll\sqrt{\frac{\hbar}{8\omz mW^2}}.
\end{equation}
In the situation considered above with $W=5\unit{\mu m}$ and
$\omz=2\pi\times 200\unit{kHz}$ this means $\epsilon_p\ll 1/400$
for \cafp\ and $\epsilon_p\ll 1/700$ for \bap. Fulfilling these
criteria is not very easy, but fortunately, the undesired
phase-difference can be cancelled by using a type of spin-echo
technique~\cite{Allen-Eberly}. Instead of generating the full
effective phase-shift of $\pi$ in a single operation, the
gate-operation can be performed in four steps: (1) Run the gate at
half the intensity, to get an effective phase-shift of $\pi/2$ on
\upup. (2) Swap the population between \down\ and \up\ by applying
single-qubit $\pi$-pulses to both ions. (3) Same as (1). (4) Same
as (2). The trick here is that the undesired phase-differences due
to a polarization error, which are obtained in step (1) and (3),
are of the same magnitude but have opposite signs (due to the
population swapping) and therefore cancel out. The gate operations
in (1) and (3) both give an effective phase-shift of $\pi/2$, even
though the population is swapped in (2), because
$\phi_\pm(\downarrow\downarrow,T)=\phi_\pm(\uparrow\uparrow,T)$
and
$\phi_\pm(\downarrow\uparrow,T)=\phi_\pm(\uparrow\downarrow,T)$.
The final $\pi$-pulse just swaps the population back. Using this
trick the gate time is doubled (neglecting the duration of the
relatively fast $\pi$-pulses), while the required intensity is
halved. Since the required intensity is proportional to $n^{-1}$
and the gate time $T\propto n$, these parameters can be
re-adjusted if an appropriate value for $n$ is available.

Finally, the imbalance also gives rise to errors in the gate
operation, which the spin-echo trick does not cancel. These errors
are of order $\epsilon_p^2$ or $\epsilon_p/n$ and hence they are
suppressed to the $10^{-4}$ level at a small but realistic value
of $\epsilon_p\sim 1\%$.
\subsubsection{Timing errors}
In case the gate time differs from the duration of a full number
of polarization rotation periods, an undesired phase-difference
will again build up. Assuming $T=\delta T+2\pi(n-1)/\Omega$ and
$\Omega \delta T\ll1$, the phase-difference is equal to
$\Delta\phi$ above, with the replacement $\epsilon_p\mapsto\delta
T^2\Omega/(2T)$. Again using $W=5\unit{\mu m}$ and
$\omz=2\pi\times 200\unit{kHz}$ and considering \cafp\ with
$n=15$, Eq.~\eqref{eq6:bad-polarization condition} translates to
$\delta T\ll 0.5\unit{\mu s}$, with a more relaxed limit for
larger $n$ (the limit is proportional to $\sqrt{n}$). Applying
electro optic modulators to control the laser pulse length, the
condition on $\delta T$ is not very severe since switching times
of a few nanoseconds can be obtained.

\subsubsection{Power fluctuations}
If the total laser power fluctuates at a frequency $\omsub{f}$,
such that the total intensity is given by
$I_0(t)=I_0[1+\epsilon_f\sin(\omsub{f}t)]$, then the resulting
fluctuations in the dipole potential integrated over the gate time
$T$ will give rise to a phase difference between \down\ and \up.
If the intensity fluctuations are random, they can in general not
be expected to cancel using the spin-echo trick. When
$\omsub{f}\sim\Omega$ the phase difference
$\Delta\phi\sim\epsilon_f I(z)(\psi_+-\psi_-)/2$, i.e., the same
phase-difference as above, just with $\epsilon_p$ replaced by
$\epsilon_f$, which means that $\epsilon_f\ll1/400$ is required
for \cafp. When $\omsub{f}\ll\Omega$ the phase-difference is
smaller by a factor of $2\pi n$ and even smaller if
$\omsub{f}\gg\Omega$. Intensity stabilization fulfilling
$\epsilon_f\ll1/400$ is not unrealistic, in fact a commercially
available Laser Power Controller already offers a power-stability
of $3\cdot10^{-4}$ within certain limits~\endnote{Laser power
controller from Brockton Electro Optics,
\texttt{www.brocktoneo.com}.}.
\subsubsection{Position fluctuations}
Fluctuations in the position of the dipole force inducing beam
give rise to intensity fluctuations, which lead to imperfect
cancellation of the Stark-shift induced phase and an unwanted
variation in the dipole force exerted on the ions. Hence, for high
quality gate operations with errors at the $10^{-4}$ level,
position jitter has to be of the order of $1\%$ of the beam waist
$W$ or smaller. This condition is most restrictive for the
situation shown in Fig~\ref{fig6:beams-ions}(c), where a pointing
stability of about 50 nm is needed when $W=5\unit{\mu m}$. Though
small, this type of stability is technically possible.
\subsubsection{Frequency fluctuations}
As for the power and position fluctuations, laser frequency
fluctuations will lead to fluctuations in the dipole potential and
hence to an imperfect cancellation of the Stark-shift induced
phase. Since laser frequencies can be very accurately controlled
and frequency fluctuations anyway are expected to be small, as
compared to the detuning from any of the two fine-structure
levels, this is not expected to play any significant role.
\section{Discussion}\label{sec:conclusion}
The gate proposal presented above has some similarities with the
gate recently demonstrated by the NIST
group~\cite{Leibfried-CNOT}; in fact the mechanism which gives
rise to the desired phase shift is exactly the same. There are,
however, also some essential differences between the two schemes.

In the NIST experiment, a dipole force along a given trap axis was
provided by the intensity-gradient of a moving standing wave
light-field. The wavelength of the light-field and hence the
period of the standing wave is set by the requirement that the
Stark-shift induced phase shift of the two qubit levels should be
zero (corresponding to $\psi_++\psi_-=0$ in our case), which is
fulfilled if the laser is tuned in between two fine-structure
levels. This can, however, give rise to a significant amount of
scattering events, as we also saw above. Even though this scheme
relies on a moving standing wave, the instantaneous force on the
two ions is required to be equal. Consequently, the ions have to
be well localized which requires cooling to the Lamb-Dicke limit
with respect to the wavelength of the light-field, which in this
case also is the Lamb-Dicke limit for the qubit operations. An
equal dipole force on the two ions was obtained by adjusting the
distance between them to an integer number of standing wave
periods, which may be difficult to generalize to perform a gate
between any two ions in a multi-ion string.

In the proposal presented here, where the dipole force is provided
by a variation in the beam-profile, the excursion of the ions from
their equilibrium position should only be smaller than the beam
waist, which is adjustable, but typically up to ten times larger
than a relevant transition wavelength. This means that except for
very tightly focussed beams the Lamb-Dicke limit criterion need
not be fulfilled and, specifically, the ions need not be cooled to
the motional ground state. Since the dipole force inducing beam
propagates perpendicular to the ion-string in our proposal,
addressing of specific ions for implementation of gates in a
multi-ion string should be feasible. A theoretical description of
this situation should also be quite straightforward by
generalizing Eq.~\eqref{eq6:wavefunction}. Finally, owing to the
polarization rotation method, the dipole force inducing laser is
allowed to be far-off resonant with respect to the relevant
internal transitions, such that a scattering probability below the
asymptotic threshold value required for fault-tolerant quantum
computation~\cite{Steane-errorcorrectionII} in principle can be
obtained in the long wavelength limit.

It should be mentioned, that the idea of using optical dipole
forces for implementing a Controlled-Z gate has also been
considered by Sasura and Steane for an array of microscopic ion
traps with a single ion in each trap~\cite{Sasura-Steane}. In many
respects this system is very similar to the one considered here
and naturally many of the considerations are the same as those
made above.

In a very recent proposal, Garcia-Ripoll, Zoller and Cirac present
a geometric gate, where the momenta of the involved ions are
controlled by absorption of photons from a discrete set of laser
pulses~\cite{Garcia-Ripoll-gate}. In this case requirements on the
pulses naturally arise for having zero displacement and for
obtaining the desired phase shift. It can be shown that these
requirements are discrete versions of those expressed through
$\tilde{g}(\omega)$ above.

In conclusion, we have presented a proposal for a geometric
quantum gate and shown that an experimental realization using the
attractive alkaline earth ions \cafp\ and \bap\ is feasible with
gate times below 100\unit{\mu s} and errors at the $10^{-4}$ level
as required for fault-tolerant quantum
computation~\cite{Steane-errorcorrectionII}.

\end{document}